%% file: main_hexalab.tex
\documentclass[5p,twocolumn]{elsarticle}

\usepackage[utf8]{inputenc}
\usepackage{graphicx}
\usepackage{url}
\usepackage{hyperref}
\usepackage{listings}

\usepackage{enumitem}

\usepackage{xspace}
\newcommand{\Hexalab}{HexaLab\xspace}
\newcommand{\hexalab}{HexaLab\xspace}
\newcommand{\f}{\Phi}

\newcommand{\meshs}{\ensuremath{\mathcal{M}_S}\xspace}
\newcommand{\meshv}{\ensuremath{\mathcal{M}_V}\xspace}

\journal{Journal of \LaTeX\ Templates}

\begin{document}

\begin{frontmatter}

\title{HexaLab.net: an online viewer for hexahedral meshes}

\author{Matteo Bracci\fnref{isti}\corref{firstauth}}
\author{Marco Tarini\fnref{unimi}\corref{firstauth}}
\author{Nico Pietroni\fnref{uts}}
\author{Marco Livesu\fnref{imati}}
\author{Paolo Cignoni\fnref{isti}}

\fntext[isti]{CNR ISTI, Pisa, Italy}
\fntext[unimi]{Università degli studi di Milano, Italy}
\fntext[uts]{University of Technology, Sydney, Australia}
\fntext[imati]{CNR IMATI, Genoa, Italy}

\cortext[firstauth]{Joint first author}




\begin{abstract}
We introduce \Hexalab: a WebGL application for real time visualization, exploration and assessment of hexahedral meshes. \Hexalab can be used by simply opening \href{http://www.hexalab.net}{www.hexalab.net}. Our visualization tool targets both users and scholars. 
Practitioners who employ hexmeshes for Finite Element Analysis, can readily check mesh quality and assess its usability for simulation.
Researchers involved in mesh generation may use \hexalab to perform a detailed analysis of the mesh structure, isolating weak points and testing new solutions to improve on the state of the art and generate high quality images. To this end, we support a wide variety of visualization and volume inspection tools. Our system offers also immediate access to a  repository containing all the publicly available meshes produced with the most recent techniques for hexmesh generation. We believe \Hexalab, providing a common tool for visualizing, assessing and distributing results, will push forward the recent strive for replicability in our scientific community.
\end{abstract}


\end{frontmatter}


\input{intro}

\input{state}

\input{overview}

\input{repository}

\input{rendering}
\input{filters}

\input{quality_vis}

\input{structure_vis}

\input{quality_measures}

\input{status}
\input{impl}

\input{end}

\section*{Acknowledgments}
This work was partially funded by the EU ERC Advanced Grant
CHANGE, agreement No 694515, and by Italian MIUR, project
``DSURF'' (PRIN 2015B8TRFM).

\bibliography{biblio}

\end{document}

%% file: intro.tex

\section{Introduction}
\label{sec:intro}

Hex-meshes, i.e.\ volumetric meshes composed of hexahedral cells, are one the of most used 3D representations for
numerical simulation, most notably by Finite Element Analysis (FEA). Application domains include
structural mechanics, heat, electricity transfer problems and simulation of other physical phenomena.
In order to be usable in a simulation,
a hex-mesh must fulfill a number of requirements, both hard and soft in nature. In other terms, it must have a sufficiently high ``quality''.

The construction of simulation grade hex-meshes for a given shape is a long standing, extremely arduous problem, which has continuously attracted interest from industry and fuels a constant (and still ongoing) research effort by more than one scientific community (Sec.~\ref{2dot1}). The faced tasks are cast, for example, as the automatic and reliable construction of a hex-mesh of an object given its surface (hex-meshing); the ``clean up'' of a given hex-mesh form pathological configurations that prevent usability (untangling); or the conversion of a tetrahedral mesh into a hex-mesh (hex-remeshing).

We introduce \emph{\hexalab}, an Open-Source software tool designed to help researchers and practitioners. \Hexalab is both an advanced hex-meshes visualizer, and a portal to an online database of results produced with existing techniques.
It is provided as a Web Application, and can be used by simply connecting to \url{www.hexalab.net}.

\begin{figure}[h]
	\centering
\includegraphics[width =1.0\linewidth]{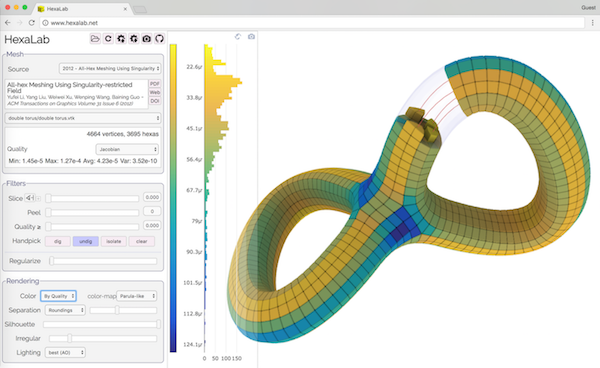}%
\caption{A screen-shot of \hexalab running inside Chrome under macOS. Model courtesy of \cite{Li:2012}.}
	\label{fig:interface}
\end{figure}

\subsection{Intended usages}
 
\noindent\textbf{Visualization:} the main function of \hexalab is to serve as an advanced 3D visualizer for hex-meshes, for the purpose of providing a visual insight on the inspected models, and thus, indirectly, on the algorithms that produced them. The visualization of a hex-mesh is complicated by the resolution of the model and, even more so, by the presence of non-trivial 3D \emph{internal} structures.
\Hexalab faces this task by offering a set of interactively controlled tools (cutting planes, etc), a shape-revealing realistic lighting (drastically improving the readability of images), novel modalities specifically designed to better communicate the shape of the cells, as well as colormaps and spatial glyphs, to reveal the quality, the topological characteristics, and so on.

\noindent\textbf{Assessment:} for the same purpose, \hexalab can be used to assess the quality of the inspected models, by providing interactive and automated techniques to perform numerical measurements and plotting histograms for the inspected model. A wide array of established measures are included.

\noindent\textbf{Presentation:} although they are produced in real-time in the context of an interactive application,  images produced by \hexalab have publication-grade quality and are intended to be used for presentation and dissemination, e.g.\ in scientific publications. The high lighting quality helps to making 3D rendering intelligible even as static (e.g.\ printed) images. 
This function relieves researchers from the tedious task of implementing \emph{ad-hoc} visualization method for the presentation of their results; 
also, such visualizations are often comparatively less effective, and are inhomogeneous across works from different authors.

\noindent\textbf{Comparisons:} \hexalab also implements mechanisms to ease comparison between hex-meshes, e.g.\ between results of competing methods over the same input, or, between ``before'' and ``after'' datasets of a hex-mesh optimization. All visualization settings can be easily shared between visualization sessions. This applies to sessions being run at the same time, for interactive side-to-side comparisons, and just as well to sessions executed much later in time. The visualization settings are also recorded in all produced images as metadata, to allow reproducibility of these images, and, thus, direct comparisons with new results in the future. Potentially, comparison is also indirectly fostered, even by results not directly dealing with each other, by their adoption of a shared visualization style.

\noindent\textbf{Batch processing:} in recent years, the number of results typically
used to evaluate remeshing algorithms increased. For this reason,
HexaLab can be used to read a collection of models, and produce
images and measurements for them in one go.

\noindent\textbf{Benchmarking:} lastly, but perhaps most crucially, \hexalab is also an easily accessible portal to a new online repository of hex-meshes, which is already fairly complete, and, we believe, is destined to grow over time with the continued usage of the tool. Direct contributions of new datasets from authors of new methods are made simple and, in our intentions, encouraged (among other things) by the resulting visibility offered by \hexalab to any such works. The original source of any work in the repository is reported to all future users.

%% file: state.tex
\begin{figure*}
\includegraphics[width =0.20\linewidth]{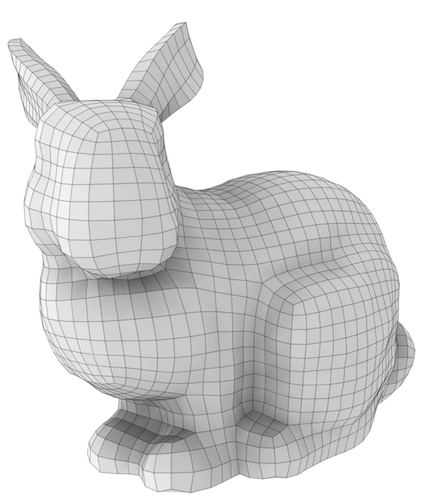}%
\includegraphics[width =0.20\linewidth]{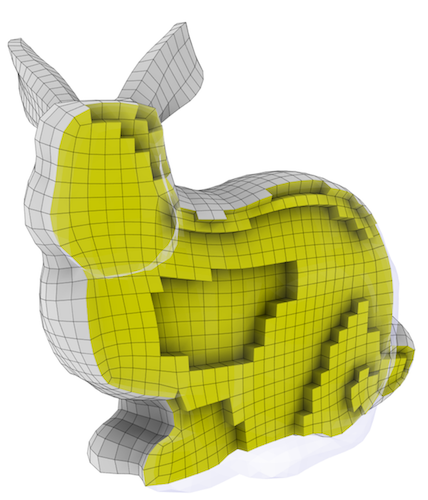}%
\includegraphics[width =0.20\linewidth]{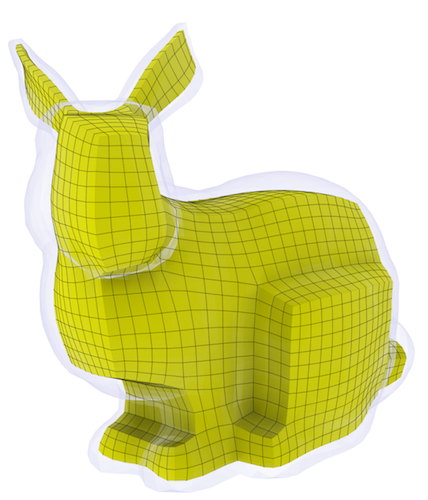}%
\includegraphics[width =0.20\linewidth]{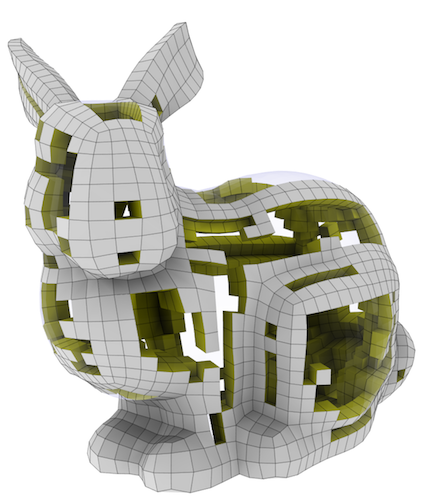}%
\includegraphics[width =0.20\linewidth]{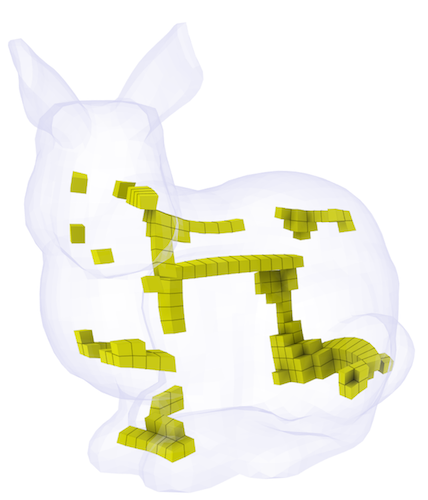}%
\caption{Examples of \hexalab filters applied to an hex-mesh to reveal its interior. From left to right: complete, unfiltered model; slicing-plane filter; peeling filter (two most external layers of cells removed); quality-thresholding filter (hexas with SJ $< 0.96$ are shown); and the combination of all these filters. ``Bunny'' model courtesy of \cite{10.1111:cgf.12959}.}
	\label{fig:clipping_plane_img}
\end{figure*}

\section{Background and Related Work}

\subsection{Hex-Mesh Generation and Processing}
\label{2dot1}
Hex-meshes are often preferred to other polyhedral based representations, such as tetrahedral meshes, because 
arguably  
they offer advantages in terms of numerical accuracy \cite{wang2004back} or, equivalently, simulation speed for the same level of accuracy (due to the smaller number of degrees of freedom). Hex-meshes generation, however, is notoriously difficult.

 In 1998 Owen \cite{owen1998survey} pointed out that hex-meshing techniques were not robust enough to be able to scale on complex shapes. Two years later, Blaker~\cite{blacker2000meeting} defined hexmesh generation as the \emph{holy grail} of meshing research. 
In the last decade, there has been a constant improvement in hex-meshing algorithms, mainly tied to advancements in volumetric parameterization techniques; yet, more research is clearly needed before fully satisfactory solutions are reached.
Here we briefly summarize only a few of the most recent advances, and point the reader to \cite{owen1998survey} for a survey on classical methods developed during 90s.
It is our opinion that availability of powerful tools to visualize, inspect, compare and analyze hex-meshes will help promoting further advance in the state of the art. 

\emph{Grid based approaches} \cite{Schneiders1996} subdivide the volume using a regular grid. Elements are aligned to the global axes,  and external vertices are snapped to the surface to better approximate the original shape.
The most recent advances in the field regard the ability to project vertices on the surface while meeting per element quality bounds \cite{Lin:2015:QGA:2828638.2828677} and the introduction of templated schemes to turn hierarchical grids (octrees) into full hex-meshes \cite{Marechal2009}. The resulting meshes often expose a complex structure, which can be simplified in post-processing \cite{gao2017simplification}\cite{gao2015hexahedral}.

\emph{Advancing front} approaches instead propagate element generation from the boundary toward the interior \cite{Kremer2014}. While these methods generate high quality elements close to the boundary surface, they tend to create badly shaped elements or leave empty voids in the interior, where different fronts meet. Alternatively, the front propagation may proceed \emph{top to bottom}. In \cite{gao2016structured} the authors propagate the front following the integral curves of a harmonic function. Front splitting/merging at the saddle points of the guiding function is also supported.

One recent trend is based on \emph{volumetric parameterizations}. The volume is first mapped on a space that admits trivial hexmeshing (e.g. via regular sampling); then the connectivity is  generated in parametric space and projected in object space using the inverse map. Mesh is eventually cured \cite{LSVT15} to remove imperfections. Common parametric spaces are cylinders \cite{LAPS17,Livesu2016} and orthogonal polyhedra (or \emph{polycubes}~\cite{tarini2004polycube}). Approaches differ to each other on how the parametric space and the map are generated. High-quality polycubes should balance map distortion with corner count \cite{fu2016efficient,huang2014}\cite{Livesu:2013:PolyCut}, achieving a good singularity alignment \cite{10.1111:cgf.12959}. As observed in \cite{Fang:2016:AMU:2897824.2925957}, polycubes can be seen as a special case of field-aligned methods \cite{CGF:CGF2014,Li:2012,jiang2014frame,kowalski2016smoothness}; the latter are more general, having internal singularities to improve the element shapes. In field-aligned methods, the mesh is generated by tracing integer iso-lines \cite{lyon2016hexex} of a volumetric parameterization aligned with a frame-field \cite{ray2016practical,huang2011boundary}. The structure and quality of the mesh depend on the singularities and the smoothness of the guiding field, respectively. Unfortunately, not all the possible field singularities are compatible with hex-meshing, and a number of heuristics have been proposed to remove incompatible configurations \cite{Li:2012,jiang2014frame,suggestedByRev1:2018}. The robust generation of fields that admit a valid hex-mesh is an open problem. Alternatively, recent field-aligned methods targeting the construction of hex-dominant meshes \cite{hybridHexa,Sokolov:2016:HM:2965650.2930662} can bypass this problem.

\subsection{Hex-Mesh Software Tools}

Alongside professional and well-established open tools for solid mesh generation and processing, such as Gmsh~\cite{geuzaine2009gmsh} and ParaView~\cite{ayachit2015paraview}, there exists a variety of smaller tools being developed by researchers operating in the field and released to the community \cite{zheng1995feview}. We give here a non comprehensive list of hex-mesh related software, considering both libraries and desktop applications.\\

A number of freely available libraries offer data structures \cite{geogram}\cite{openvolume}\cite{cinolib} to import/export a hex-mesh and process it. These tools usually have a dedicated visualization front-end (e.g., Graphite~\cite{graphite} is the visual front-end of GeoGram~\cite{geogram}), but the functionalities they offer are more limited if compared to the ones we offer in \hexalab (e.g., no histograms, no advanced lighting, no advanced inspection tools for the mesh interior and structure).

Other software tools focus on mesh synthesis and processing, such as Hexotic (which implements \cite{Marechal2009} and is distributed by Distene within their MeshGems), LibHexEx~\cite{lyon2016hexex} (which enables the extraction of a hexmesh starting from an integer grid map) or Mesquite~\cite{brewer2003mesquite} (which serves to maximize per element quality). These tools are orthogonal to us, as they focus on the processing of a hex-mesh but do not offer any visualization facility whatsoever.

Summarizing, our positioning is in-between professional tools and tools being developed and maintained by pure researchers: we offer high quality visualization and analysis tools which are better than the ones provided by the rest of the research community, at the expense of a simplified setup (we run on browser) which is much lighter than the one usually required by professional software.

\subsection{3D Online Visualization}
Compared to desktop applications, web-applications are by design lightweight and have many several desirable 
characteristics, such as extreme portability (being based on a natively cross-platform technology), immediate availability to users (due to the absence of an installation phase), safety (due to browsers being protected environments), amenability to online and distributed applications (due to the inherent client/web-server setup), 
and easiness of deployment and maintenance (due to instant updates).

Traditionally, high-quality real-time 3D rendering of complex scenes has been considered a computationally demanding task, requiring GPU support, and therefore advanced visualization tools have been limited to desktop applications. 
In the last years, the arise of a widely supported web-based 3D API, WebGL by Khronos, fueled an increase of web-based 3D visualization tools. 

Many of these tools target specific domains and are to use, embedding a small set of advanced specialized functionalities that are simple to understand. 
For example, in the field of molecular visualization, after the first Java-based approaches, like JMol\cite{hanson2010jmol}, a number of small specialized visualization tools have been proposed \cite{callieri2010visualization,rose2015ngl,li2014iview}.
Similarly, various web-based applications have been developed for volume visualizations tasks \cite{garcia2016volume,Congote:2011,polys2012cross}, confirming the appeal of specialized, lightweight visualization tools. 

General-purpose tools to display collections of 3D models have also been one of the most common tasks for web-based visualization applications, such as Sketchfab or 3DHop \cite{Potenziani2015}, among many others; we refer the reader to \cite{SCDPP17} for a discussion of the possibilities and the issues of delivering 3D collections on the Web.

%% file: overview.tex
\section{Overview}
\label{sec:overview}

\Hexalab can be used to visualize and assess the quality of hex-meshes either provided by the user in standard formats 
or downloaded from its own online integrated library of reference hex-meshes from recent literature (Sec.~\ref{sec:repository}).

The visualization and exploration process is based on a very simple approach: in a trackball-controlled, real-time, high quality 3D rendering of the inspected mesh (Sec.~\ref{sec:rendering}) the user can interactively apply on it three different kinds of \emph{exploration tools}:

\begin{description}
\item[Cell Filtering Tools:] (Sec.~\ref{sec:filters}) that allow users to hide portions of the hexahedral mesh so to expose internal structures which would otherwise be occluded. Criteria to hide cell can involve the use of a clipping plane, distance from mesh boundary, geometric quality, and handpicking of specific cells;

\item[Quality Assessment Tools:] (Sec.~\ref{sec:quality}) that allow users to visually and numerically assess the geometric quality of the hexahedral cells, both individually (via color-coding), and as aggregated measures (through histograms and statistics), according to a number of standard quality measure (Sec.~\ref{sec:measures}). Inverted (and concave) cells are treated and highlighted separately;

\item[Global-structure Visualization Tools:] (Sec.~\ref{sec:structure}) that depict the configuration of irregular elements (especially edges) across the entire mesh under inspection, thus revealing its global topological structure.
\end{description}


\Hexalab offers visualization status management (Sec.~\ref{sec:status}):
at any moment, the current status of the interactive visualization can be readily captured, stored, and reused across visualization sessions. The status includes the setting of every active tool, the trackball-determined view-direction, any rendering parameters, the selected quality measure etc.  This achieves reproducibility of any previously obtained image and allows visual comparisons, under the exact same conditions, of different datasets (e.g. for comparing alternative meshings of the same object).

While in this work we do not focus on interfaces, we strive to complete \hexalab with a reasonably intuitive Graphic User Interface (GUI). GUI elements will be described in the following sections, contextually with the mechanisms they control.

%% file: repository.tex

\begin{figure*}[!h]
	\centering
	\begin{tabular}{@{}c@{}c@{}c@{}c@{}}
		\includegraphics[width =0.25\linewidth]{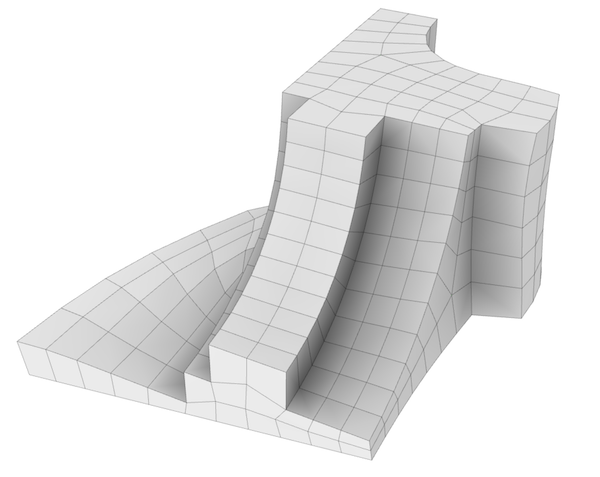}&
		\includegraphics[width =0.25\linewidth]{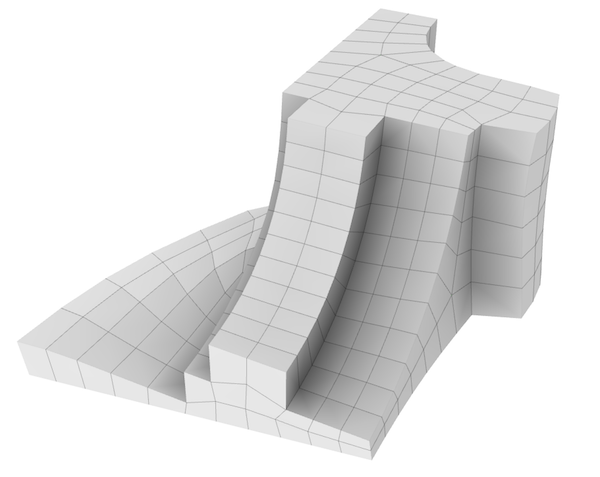}&
		\includegraphics[width =0.25\linewidth]{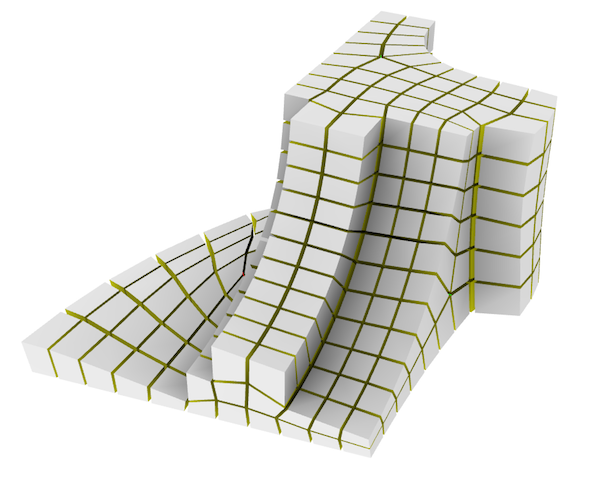}&
		\includegraphics[width =0.25\linewidth]{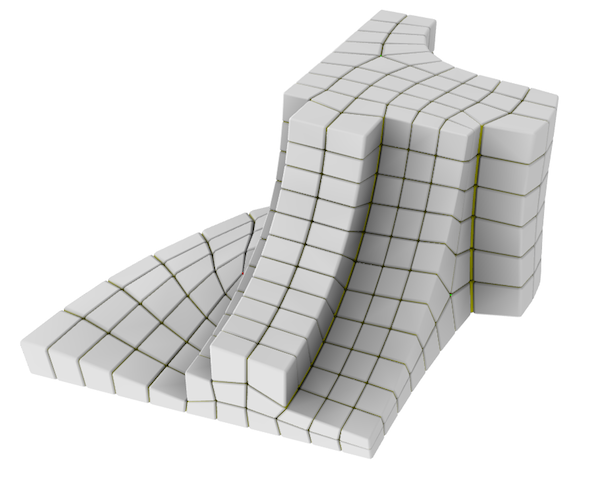}\\
		\includegraphics[width =0.25\linewidth]{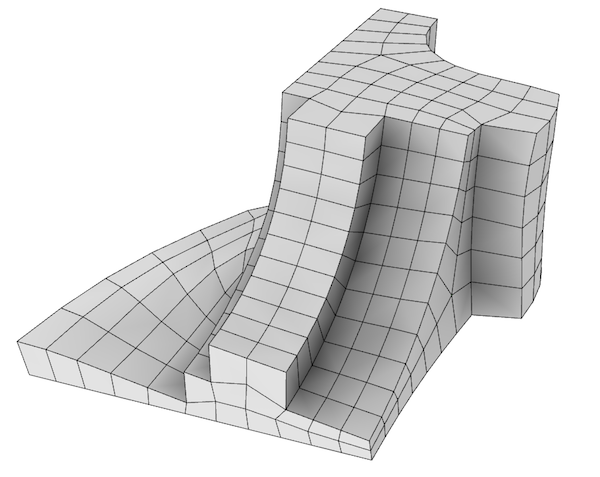}&
		\includegraphics[width =0.25\linewidth]{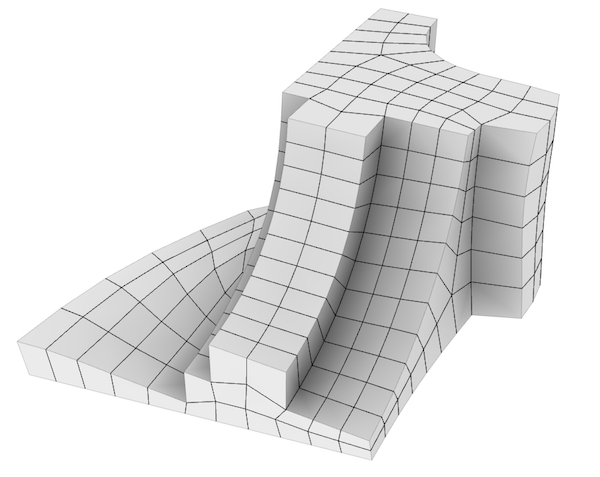}&
		\includegraphics[width =0.25\linewidth]{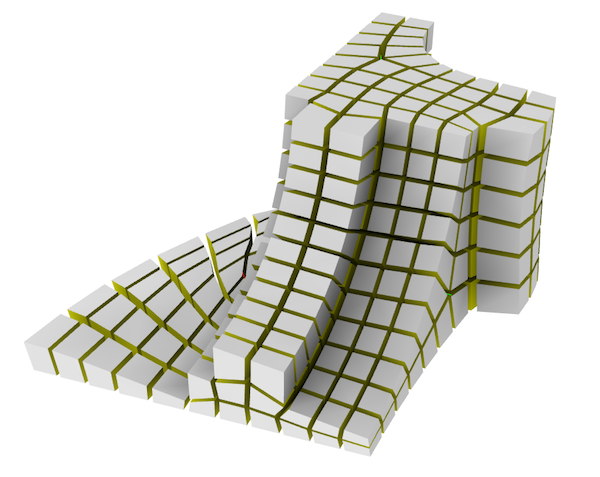}&
		\includegraphics[width =0.25\linewidth]{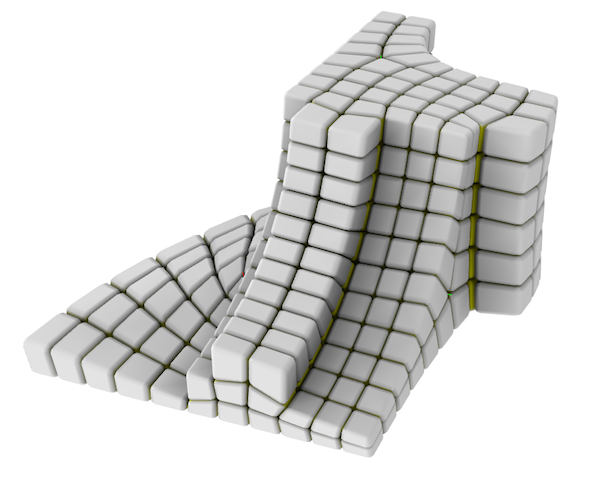}\\
		(a) & (b) & (c) & (d) \\
	\end{tabular}
	\caption{(a) the traditional rendering of \meshs to represent \meshv. (b) darkness-coded mode. (c) crack mode. (d) rounded mode. Bottom row: the same modes, for a different selection of darkness$/$width$/$radius. Model courtesy of \cite{Li:2012}.}
	\label{fig:separationModes}
\end{figure*}

\section{Online mesh repository}
\label{sec:repository}

\Hexalab doubles as an easily accessible, online repository of the publicly available hex-meshes produced by different techniques appeared in recent literature. 
Currently, the repository consists of a total of 272 hex-meshes, from 15 different papers published in the last 7 years: \cite{Gregson:2011,Li:2012,Livesu:2013:PolyCut,huang2014,Livesu:2015,Fang:2016:AMU:2897824.2925957,10.1111:cgf.12959,Ming2016,Livesu2016,Gao16,Shang2017,Wu:2017,LAPS17,Wu:2018,newDataset}. 

This collection of hex-meshes is stored, together with the sources, on the git repository of \hexalab; it is made available directly by the \hexalab GUI: the user can simply invoke the visualization of any stored hex-mesh, by selecting one source, and then one model from that source. The requested hex-mesh is then automatically downloaded from the repository, and presented to the user. 
Contextually, \hexalab fully reports the data source by providing the bibliographic reference, a link to the DOI, and when available, to the PDF and the web page of the referenced article.

The maintenance and updating of the repository, when new results will appear, or other authors will make their data available, will be done by relying on the well-known \emph{pull-request} mechanism made available by the GIT distributed version control system. 
In this way, we offer to the research community a direct and simple way of proposing additions of results to the repository, which will in turn foster comparisons against further advancements.  




%% file: rendering.tex

\begin{figure}[h]
	\centering
	\includegraphics[width =0.55\linewidth]{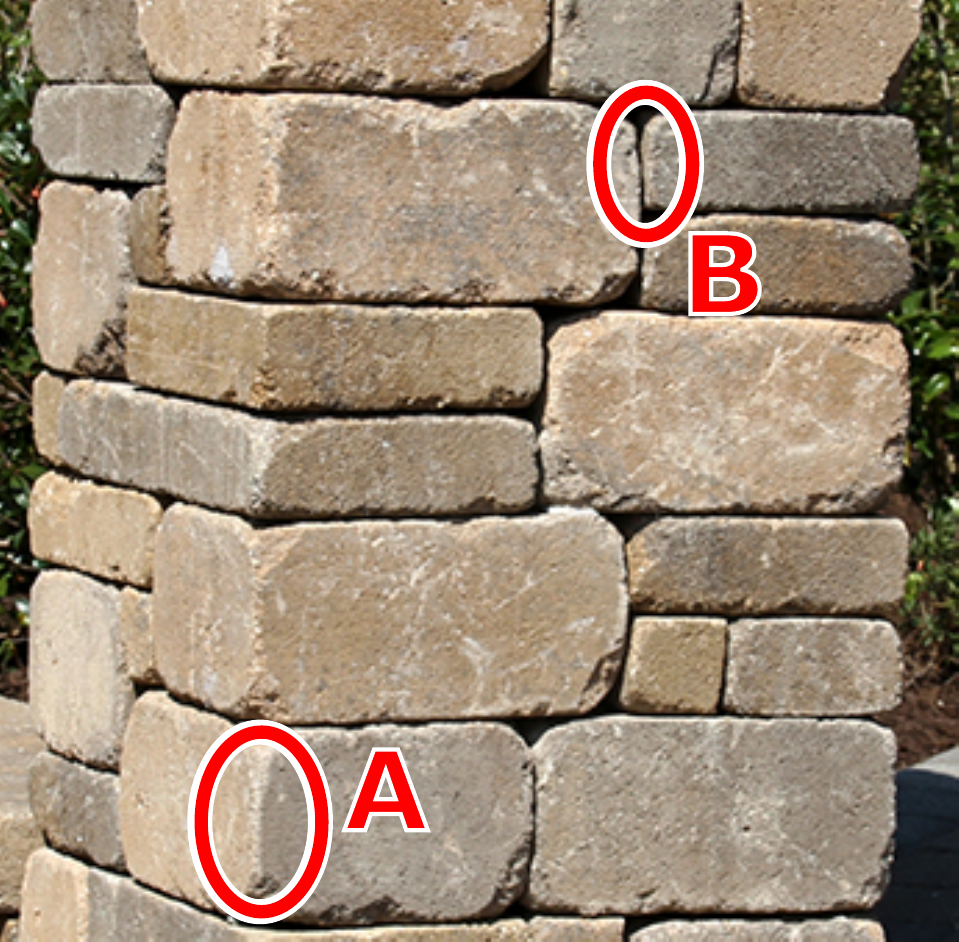}
	\caption{In this natural image of a stone brick structure, the edges that separate faces belonging to the same volumetric cell (e.g.\ in circle A) are immediately recognizable from edges that separate distinct volumetric cells (e.g.\ in circle B), although no line is explicitly added and cells are in full side-to-side contact. This inspires our Rounded mode for visualizing hex-meshes, which exploits a similar principle.}
\label{fig:brickwall}
\end{figure}

\begin{figure*}[!ht]
	\centering
\includegraphics[width=0.33\linewidth]{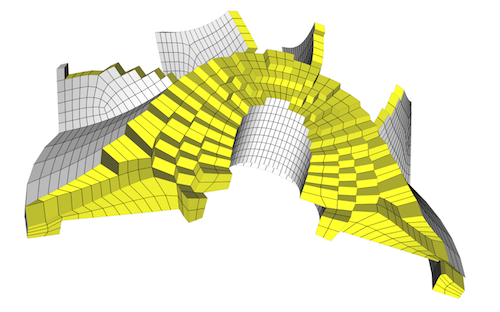}%
\includegraphics[width=0.33\linewidth]{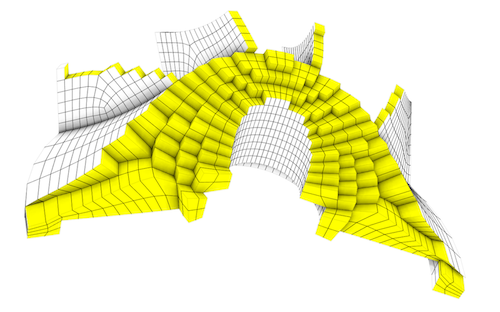}%
\includegraphics[width=0.33\linewidth]{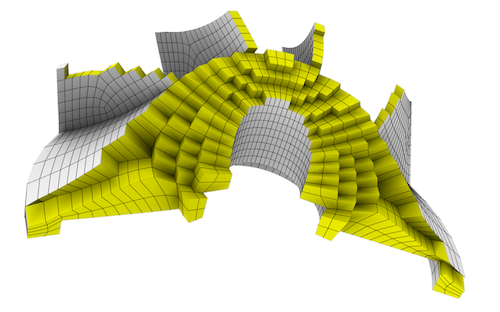}\\
\includegraphics[width=0.33\linewidth]{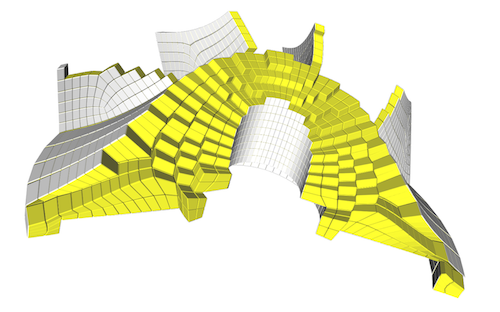}%
\includegraphics[width=0.33\linewidth]{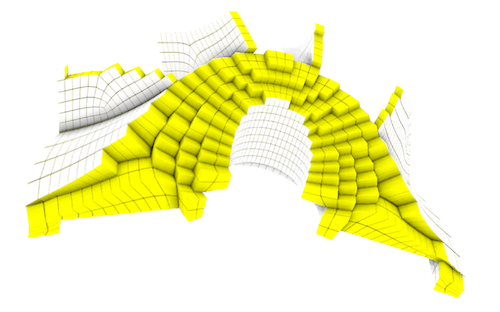}%
\includegraphics[width=0.33\linewidth]{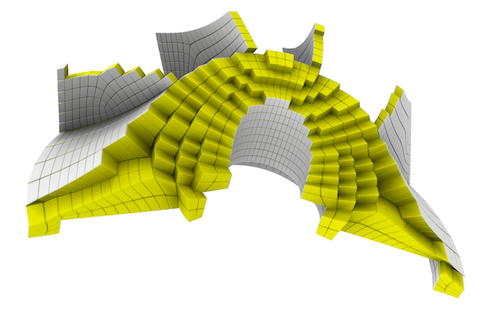}\\
	\caption{
	The lighting modes offered by \hexalab. From left to right: direct illumination only, global illumination approximated with Screen Space Ambient Occlusion (SSAO), and with Object Space Ambient Occlusion (OSAO), which is \hexalab default. Global illumination schemas are more demanding, but are more effective at convening the 3D space, especially OSAO. Bottom: the improvement is further enhanced when ``rounded mode'' is used to separate elements. Model courtesy of \cite{Li:2012}.}
	\label{fig:ao}
\end{figure*}

\section{Hexahedral Mesh Rendering}
\label{sec:rendering}

The core part of \hexalab consists in the visualization of the inspected hex-mesh as a solid object with a set of cells filtered out (see Sec.~\ref{sec:filters}), so as to reveal the interior structure.


Typically, the visualization of a polyhedral volumetric mesh \meshv renders just the boundary \meshs (a polygonal surface), using a flat shading model and with mesh edges overdrawn as line segments (see e.g. \cite{graphite}). In our case, \meshv is a pure hexahedral mesh, and \meshs is a pure quad mesh. 

We observe that this common edge rendering choice has a significant shortcoming: it is impossible to distinguish how many different cells are incident to an edge. Indeed, the edges of \meshs which separate difference cells of \meshv and the edges which just separate different faces of the \emph{same} cell of \meshv are depicted in the same way. 

A brute force approach to overcome this edge ambiguity problem appears in \cite{Sokolov:2016:HM:2965650.2930662}, and consists in drawing, instead of just the surface, the whole hex-mesh \meshv with each cell slightly smaller. This approach, that technically exposes each face and therefore the local arrangement of cells near the boundary, does not scale well in terms of performances, due to the cubic explosion of primitives to be rendered at screen.

In \hexalab we propose three visualization modes to disambiguate the surfaces representing the volumetric mesh: 
\begin{description}
\item[Darkness-coded edges:] in the wireframe, we  color the edges with thin, semi-transparent darker lines when they just separate two faces of the same cell. Opacity is higher for edges separating different cells, and progressively increased for any additional internal elements sharing such edge. In other words, the opacity of each edge on \meshs is made proportional to the number of (non-hidden) cell elements in \meshv sharing the boundary edge.
This method is undemanding in terms of resources, and greatly helps disambiguating. On the downsides, it is not self-explicative, as the mapping between line opacity and number of elements is arbitrary (see Fig.~\ref{fig:separationModes}, b). 

\item[Fissure mode:] we separate adjacent cells by a small fissure so to reveal the local inner structure of the mesh. For the sake of rendering speed, fissures are limited to cells on the boundary of \meshv and, as explained in Sec.~\ref{sec:fissureExtr}, the rest of the mesh will remain occluded. Specifically, we only separate edges and faces of \meshv which share at least one vertex on \meshs (see Fig.~\ref{fig:separationModes}, c). While this method helps solving the visual ambiguity we are addressing, it tends to clutter the screen and sometimes it is not readable.

\item[Rounded mode:] we observe that there are real world examples of cell arrangements (see Fig~\ref{fig:brickwall}) where the structure is immediately recognizable. This inspired us to define an additional visualization technique where each cell is rounded around the edge. Technically, this leaves a small elongated gap around each edge of the mesh. Similarly to the above mode, we only render the surface of this gap in the immediate proximity of the surface. More details are found in Section~\ref{sec:round_mesh}. We found this method to produce the most readable images (see Fig.~\ref{fig:separationModes}, d). 
On the other hand, this mode is a bit more resource demanding (the number of rendering primitives increases by an average factor of 12), and it can be argued that this adds geometric features (the rounded corners) which are non-existing in the input dataset.
\end{description}



Visualization modes are offered to the user as three alternatives, selectable via a combo-box.
The three modes can be further customized by editing one single parameter, which determines respectively: an opacity multiplier for the Darkness-coded edges mode; the gap size for Fissure mode; and round radius for Rounded mode (Fig.~\ref{fig:separationModes}, bottom). Since the meaning of these parameters is similar across all the modes, we avoid cluttering the interface and present the user a single slider to choose the parameter which is appropriate for the currently selected mode.


\subsection{Lighting and Shading} 

It has been observed several times \cite{Langer99perceptionof,Tarini2006} that a global lighting model is extremely helpful to facilitate a thorough comprehension of the shape of a 3D object. This is especially true when the depicted shapes are not necessarily familiar to the observer, as it is the case for hex-meshes under arbitrary filters. The problem is further exacerbated for static images (for example, in a scientific article) where the observer cannot rely on interactivity to disambiguate. 
For this reason, we consider rendering hex-meshes with only a local direct illumination inadequate, 
and we employ a view-independent Ambient Occlusion (AO) term \cite{Mendez2009,Landis02,Zhukov98} that can be pre-computed and stored at vertices of \meshs (Fig.~\ref{fig:ao}). 
%
The computation of the AO terms happens automatically in background whenever is needed, and takes only a few seconds to complete (less than 5s for the most complex model in the database, on a 2012 MacBook Pro). For the sake of interactivity, in the short time prior completion a screen-space approximation is used \cite{Bavoil2008,Mittring2007,Ritschel2009} as a temporary fall-back mode. This allows users, for example, to interactively sweep the slicing plane through the volume while still seeing a comprehensible representation.





When the color is not used to map element quality (see Sec.~\ref{sec:quality}), \hexalab defaults a simple yellow--white  color scheme to  differentiate faces of \meshs found on original mesh boundary of the mesh (prior to any filtering operation) from interior faces.

%% file: filters.tex

 \begin{figure*}[t]
\hspace{-0.35cm}\includegraphics[width =0.262\linewidth]{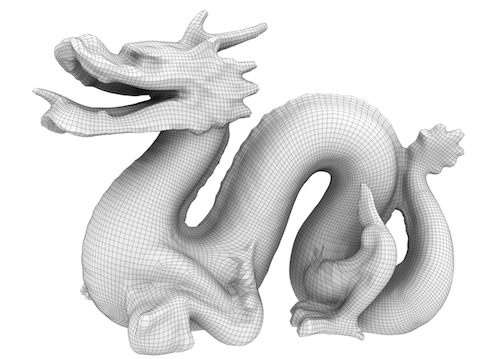}%
\hspace{-0.35cm}\includegraphics[width =0.262\linewidth]{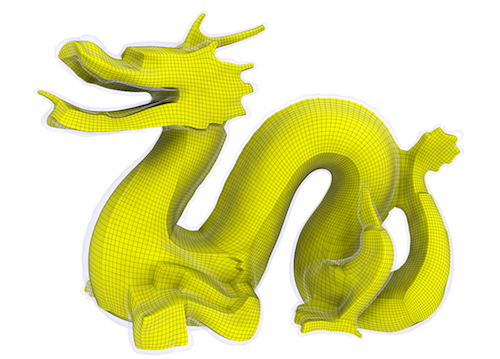}%
\hspace{-0.35cm}\includegraphics[width =0.262\linewidth]{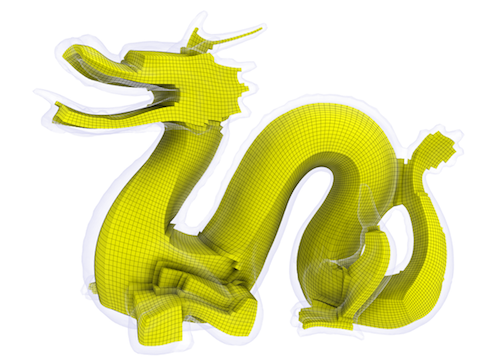}%
\hspace{-0.35cm}\includegraphics[width =0.262\linewidth]{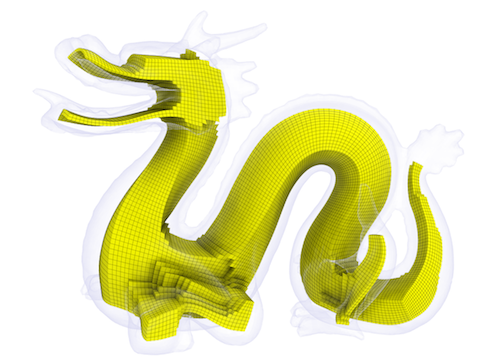}\\[2ex]\noindent%
\hspace{-0.35cm}\includegraphics[width =0.262\linewidth]{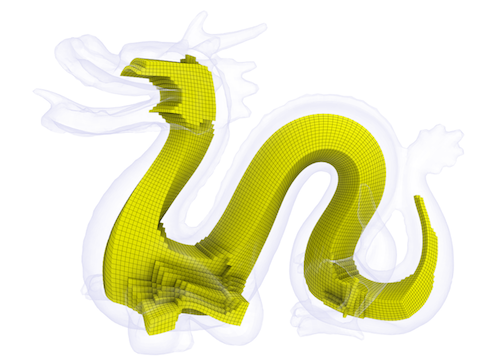}%
\hspace{-0.35cm}\includegraphics[width =0.262\linewidth]{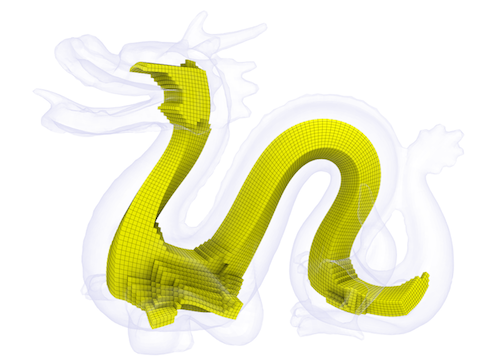}%
\hspace{-0.35cm}\includegraphics[width =0.262\linewidth]{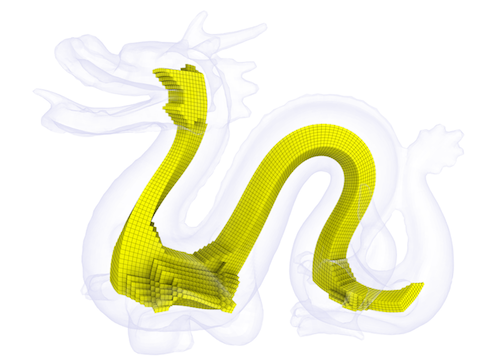}%
\hspace{-0.35cm}\includegraphics[width =0.262\linewidth]{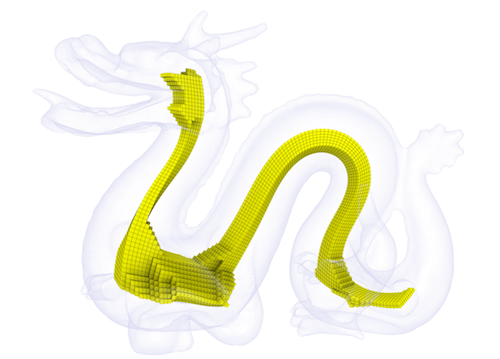}%
\caption{Several layers of peeling reveal hexahedral structure of the dragon follows the overall shape of the dragon. Model courtesy of \cite{huang2014}.}
	\label{fig:peeling}
\end{figure*}

\begin{figure}[ht]
\hspace{-0.9cm}\includegraphics[width =1.1\linewidth]{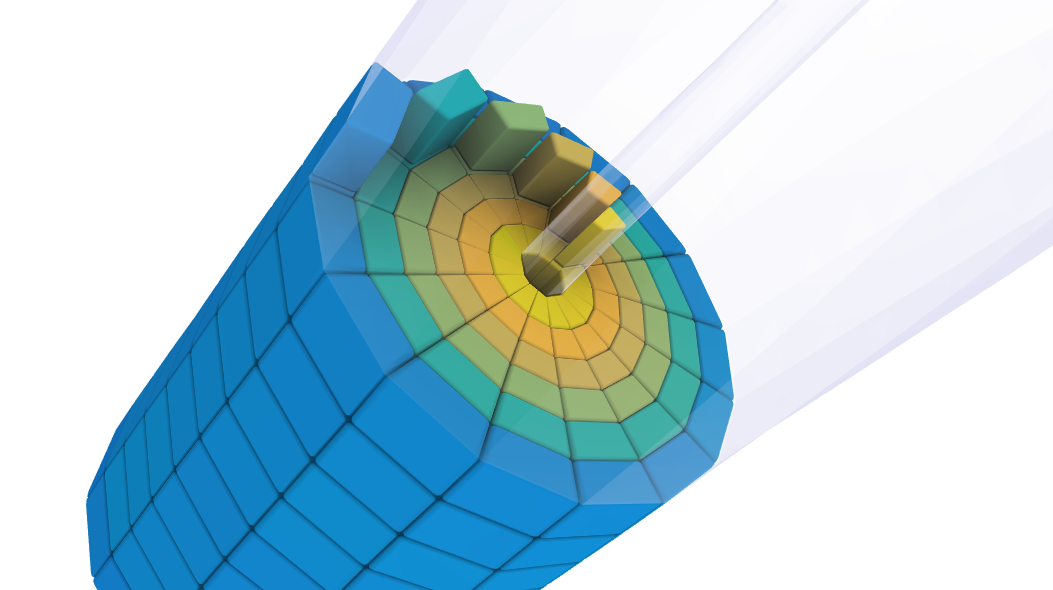}
	\caption{Sometimes, revealing ({\em undigging}) a few hidden cells produce an image which communicates the structure of the meshing in a very intuitive way. Model courtesy of \cite{LAPS17}.}
	\label{fig:undigging}
\end{figure}

\begin{figure*}
\includegraphics[width =0.25\linewidth]{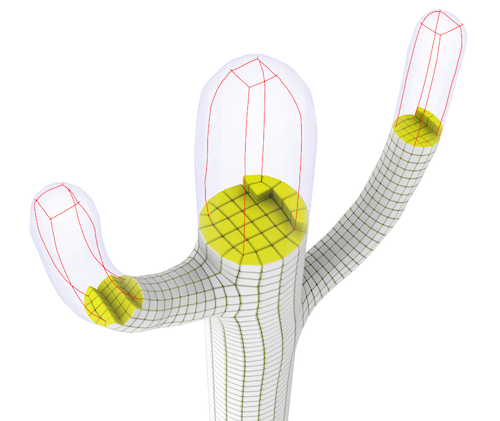}%
\includegraphics[width =0.25\linewidth]{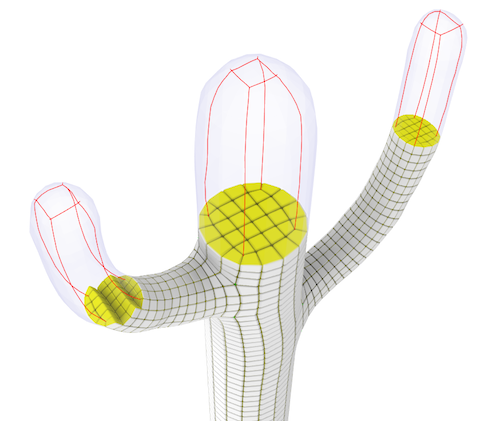}%
\includegraphics[width =0.25\linewidth]{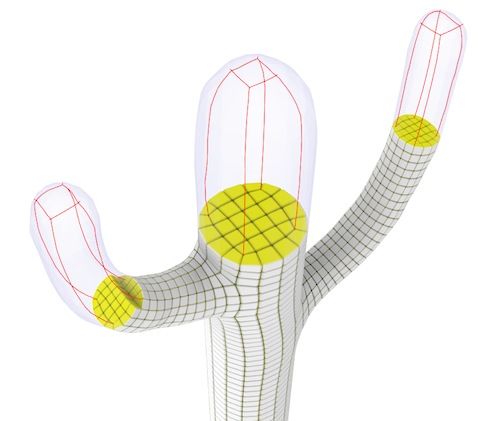}%
\includegraphics[width =0.25\linewidth]{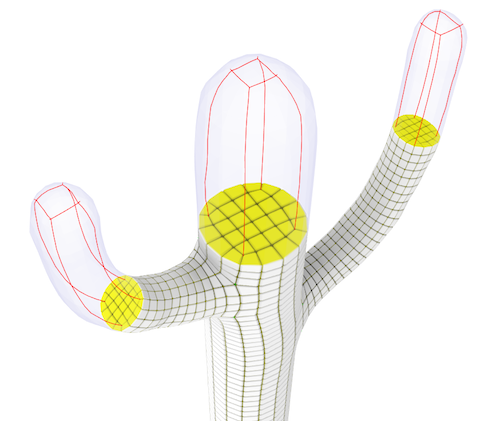}
	\caption{An example of regularization filter in action, with strengths ranging from 0 (left) to 4 (right): the boundary of the set of the cells removed by the slicing plane becomes increasingly clean. Model courtesy of \cite{Livesu2016}.}
	\label{fig:regularization}
\end{figure*}

\section{Cell filters}
\label{sec:filters}

\Hexalab provides four ways to filter away elements.
Filters all work in a consistent way, that is, by flagging certain cells as hidden, and temporarily removing them from the visualization. Filters differ to each other only on the way they flag elements, and can be used either alone or through any combination of them (Fig.~\ref{fig:clipping_plane_img}).

\begin{description}
\item[Slicing plane:] any cell whose barycenter falls behind a user specified ``slicing plane'' is filtered out from the view. In other words, the mesh is intersected with a half-space bounded by the slicing plane;

\item[Peeling:] any cell with a hop distance from the original mesh boundary smaller than a user-specified ``minimal depth'' value is filtered away (Fig.~\ref{fig:peeling}); the hop distance is computed using face-to-face adjacency. In other words, a number equal to the ``minimal depth'' of successive peels are removed in succession from the original mesh, A peel is defined as the union of the cell elements with an exposed face (Fig.~\ref{fig:peeling});

\item[Quality Thresholding:] any cell with a quality not worse than a user-specified ``quality'' is removed. 
This is useful to isolate and highlight the problematic regions, hiding the good ones;

\item[Manual Selection:] any selection obtained with the previous filters can be manually tweaked by using two operations: addition and removal of manually selected elements. These operations are triggered by pointing on a quad face $f$ on the boundary of the currently displayed mesh: 
the ``dig'' tool hides the non-hidden cell that shares $f$; the ``undig'' tool reveals the hidden cell that shares $f$ (if it exists). In both cases, the selected cell is uniquely identified by $f$. 
Finally, the ``isolate'' tool hides all cells except the selected one (neighboring cells can then be progressively added with the ``undig'' tools);
this is intended as a way to help users discerning the structure of intricate configurations. An example is shown in Fig.~ \ref{fig:undigging}, where, by undigging a few hidden cells, it is possible to easily understand the mesh configuration.
\end{description}


\begin{figure*}[t]
	\hspace{-0.5cm}
	\begin{tabular}{@{}c@{}c@{}c@{}}
		\includegraphics[width =0.35\linewidth]{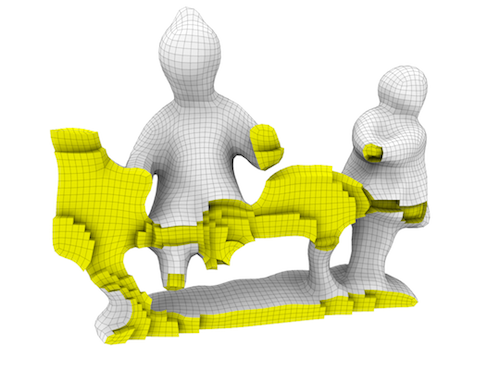}&
		\includegraphics[width =0.35\linewidth]{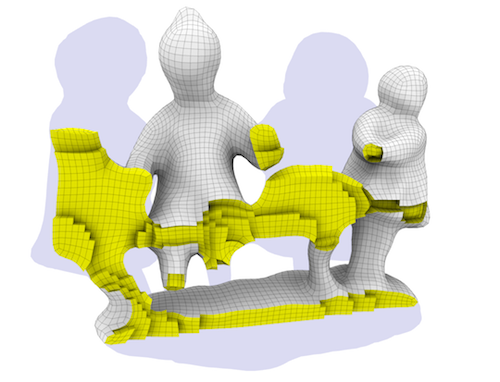}&
		\includegraphics[width =0.35\linewidth]{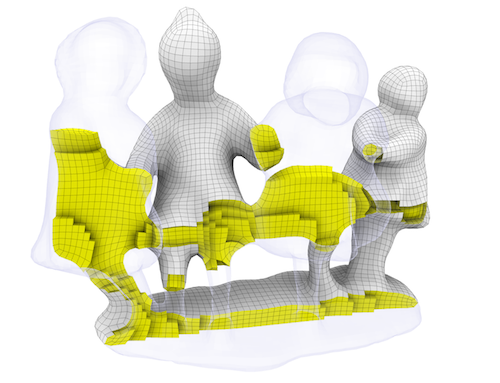}\\
	\end{tabular}
	\caption{Left: no silhouette. Middle: flat silhouette. Right: semitransparent silhouette. Model courtesy of \cite{Ming2016}.}
	\label{fig:silhouette}
\end{figure*}

\subsection{Filter regularization}
The user can elect to automatically polish the selection by increasing the regularity of the resulting boundary (i.e.\ the quad mesh)
to some extent. We implement this with a morphological approach. Specifically, the set of filtered elements is first dilated $n$ times and then eroded $n$ times, for a ``strength'' integer parameters $n$ (ranging between 0 and 5). An erosion consists in the removal from the filtered set of any cell with a vertex on the current boundary, and vice versa a dilation consists in the addition.
When this option is activated, the selection is regularized after every change of the slicing plane or the peeling tool (the other filters are unaffected). A regularized selection can help clarifying the mesh topology by revealing the structure of internal planes (see Fig.~\ref{fig:regularization}).


\subsection{Displaying filtered elements}
The elements which are filtered out are (almost entirely) omitted from the rendering, in order not to clutter the view and
not impact the readability of the currently exposed internal part.
Optionally, they can be displayed as a pale silhouette, either uniformly colored or with a very light shading, for the purpose of providing a visual hint of the spatial context, in the original mesh, of the currently visible parts (see Fig.~\ref{fig:silhouette}).

\subsection{GUI and customization} 
The user controls the visibility of the filtered elements with one slider (labeled ``silhouette''). The leftmost position completely hides filtered elements, whereas sliding to the right progressively shows the pale silhouette. By default, the slider is at the rightmost position.

The \emph{peeling} and \emph{quality thresholding} are controlled with a single slider determining the minimal depth and the maximal quality respectively.
The \emph{slicing plane} tool requires the user to identify the desired plane. 
Providing an intuitive interface for the selection of an arbitrary plane is not straightforward; we bypass this problem by leveraging the trackball, and simply providing a command (activated by a ``set plane'' button on the GUI) which uses the current view direction as the normal of the slicing plane (optionally, shift clicking this button round this vector to the closest axis-aligned direction). The facing of the plane is selected so that the cut surface faces toward the user. The offset of the plane then is controlled with a slider, analogous to the other two filters.
To increase simplicity of use, the full-scale values of the three sliders are normalized, at both ends, so that the left extreme always means a null filter (all cells are visible), and the right extreme to mean a complete filter (all cells are hidden). To this end, \hexalab silently computes, and keeps up-to-date, data for the current mesh like maximal and minimal extension along the slicing plane direction, depth of most internal element, and quality range over all the elements.

A potentially useful operation that the user might want to perform is to invert the current orientation of the slicing plane, so that the filtered portion of the mesh is  reversed, and the region of interest can be investigated ``from the other side''. This functionality can be accessed with a designated button, but can also be triggered by the set-plane button (if, upon activating it, the current view direction is detected to be close to opposite to the current slicing plane 
direction~--~within a tolerance of 20 degrees~--~then the slicing plane direction is exactly inverted). 




%% file: quality_vis.tex
\section{Cell Quality Visualization}
\label{sec:quality}

\begin{figure}[tbp]
  \centering
  \begin{tabular}{@{}c@{}}

  \begin{tabular}{@{}c@{}c@{}c@{}}%
\includegraphics[width =0.33\linewidth]{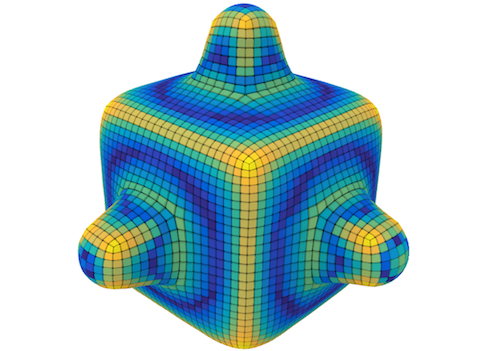}&
\includegraphics[width =0.33\linewidth]{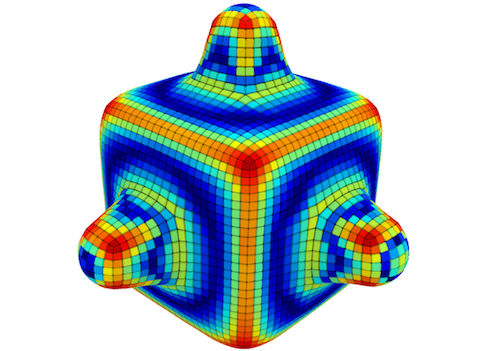}&
\includegraphics[width =0.33\linewidth]{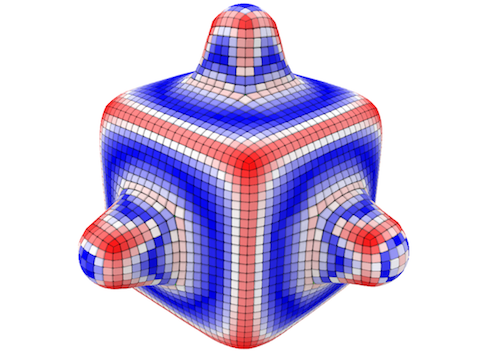}\\
  \end{tabular}\\

  \begin{tabular}{@{}c@{}c@{}c@{}}
    \includegraphics[width =0.33\linewidth]{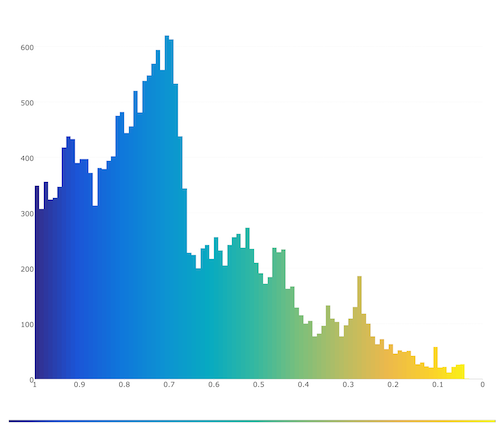}&
    \includegraphics[width =0.33\linewidth]{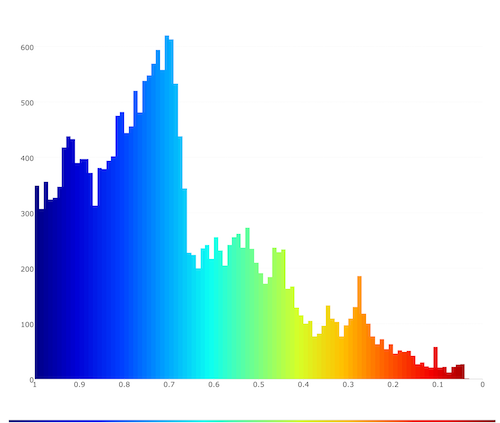}&
    \includegraphics[width =0.33\linewidth]{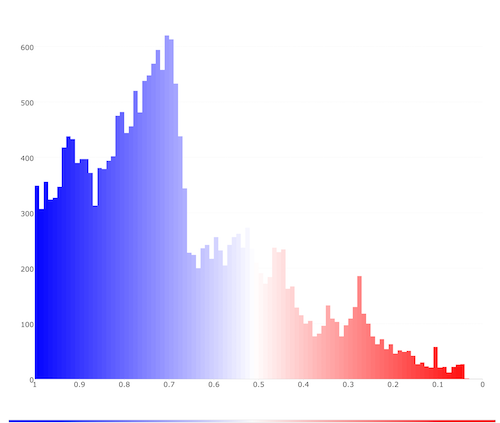}\\
    Parula & Jet & Red-Blue\\
  \end{tabular}\\

  \end{tabular}

  \caption{Different color mappings on hex quality (top), and associated histograms (bottom). Model courtesy of \cite{10.1111:cgf.12959}.}

 \label{fig:quality_filter_img}
\end{figure}

\begin{table*}[h]
\centering
\begin{scriptsize}
\begin{tabular}{| l | l | c | c | c | c | c |}
\hline
\textbf{Metric}   & \textbf{ID} & \textbf{Full}    & \textbf{Acceptable}  & \textbf{Value for}& $\f(q) $& $\f^{-1}(q) $ \\
              &             & \textbf{range} & \textbf{range}         & \textbf{unit cube}&&\\
\hline
&&&&&&\\
Diagonal               & DIA   & $[0,1]$            &  $[0.65,1]$  & 1 & $q$ & $q$ \\
&&&&&&\\
Distortion             & DIS   & $(-\infty,\infty)$ &  $[0.5,1]$  & 1& $\frac{q - q_{min}}{q_{max} - q_{min}}$ & $ q_{min} + q(q_{max} - q_{min})$\\
&&&&&&\\
Edge Ratio             & ER    & $[1,\infty)$       &  --- & 1& $\frac{q_{max} - q}{q_{max} - 1}$ & $1 + (1-q)(q_{max} - 1)$\\
&&&&&&\\
Jacobian               & J     & $(-\infty,\infty)$  &  $[0,\infty)$ & 1&$\frac{q - q_{min}}{q_{max} - q_{min}}$ & $ q_{min} + q(q_{max} - q_{min})$\\
&&&&&&\\
Maximum Edge Ratio     & MER   & $[1,\infty)$  &  $[1,1.3]$ & 1& $\frac{q_{max} - q}{q_{max} - 1}$ & $1 + (1-q)(q_{max} - 1)$\\
&&&&&&\\
Maximum Asp. Frobenius & MAAF  & $[1,\infty)$  &  $[1,3]$  & 1& $\frac{q_{max} - q}{q_{max} - 1}$ & $1 + (1-q)(q_{max} - 1)$\\
&&&&&&\\
Mean Asp. Frobenius    & MEAF  & $[1,\infty)$  &  $[1,3]$  & 1 & $\frac{q_{max} - q}{q_{max} - 1}$ & $1 + (1-q)(q_{max} - 1)$\\
&&&&&&\\
Oddy                   & ODD   & $[0,\infty)$  &  $[0, 0.5]$  & 0 & $\frac{q_{max} - q}{q_{max}}$ & $ (1-q)\: q_{max}$\\
&&&&&&\\
Relative Size Squared  & RSS   & $[0,1]$  &  $[0.5,1]$  & ---& $q$ & $q$\\
&&&&&&\\
Scaled Jacobian        & SJ    & $[-1,1]$ &  $[0.5,1]$  & 1& $\max(q,0)$ & $q$\\
&&&&&&\\
Shape                  & SHA   & $[0,1]$  &  $[0.3,1]$  & 1& $ q $ & $ q $\\
&&&&&&\\
Shape and Size         & SHAS  & $[0,1]$  &  $[0.2,1]$  & ---& $ q $ & $ q $\\
&&&&&&\\
Shear                  & SHE   & $[0,1]$  &  $[0.3,1]$  & 1& $ q $ & $ q $\\
&&&&&&\\
Shear and Size         & SHES  & $[0,1]$  &  $[0.2,1]$  & ---& $ q $ & $ q $\\
&&&&&&\\
Skew                   & SKE   & $[0,\infty)$  &  $[0,0.5]$  & 0& $\frac{q_{max} - q}{q_{max}}$ & $ (1-q)\: q_{max}$\\
&&&&&&\\
Stretch                & STR   & $[0,\infty)$  &  $[0.25,1]$  & 1& $\frac{q}{q_{max}}$ & $ q \: q_{max}$\\
&&&&&&\\
Taper                  & TAP   & $[0,\infty)$  &  $[0,0.5]$  & 0& $\frac{q_{max} - q}{q_{max}}$ & $ (1-q)\: q_{max}$\\
&&&&&&\\
Volume (signed) & VOL   & $(-\infty,\infty)$  &  $[0,\infty)$  & 1 & $\frac{q - q_{min}}{q_{max} - q_{min}}$ & $ q_{min} + q(q_{max} - q_{min})$\\
&&&&&&\\
\hline
\end{tabular}
\end{scriptsize}
\caption{List of per-element metrics supported in \Hexalab. To offer a consistent color-coded quality visualization, we map each metric in the normalized interval $[0,1]$, where $0$ corresponds to the worst quality and $1$ to the best quality. The functions we use to move from the native range to the normalized range ($\f$) and vice-versa ($\f^{-1}$) are shown in the two rightmost columns of the table. For unbounded metrics, $q_{max}$ and $q_{min}$ refer to the highest and lowest quality values measured in the mesh. 
For details on the computation of each metric, the reader is referred to \protect\cite{stimpson2007verdict}.}
\label{tab:quality_metrics}
\end{table*}

The quality of individual cells is important in many applications, like Finite Element Analysis where a single badly-shaped element can impair the entire simulation. 

\Hexalab allows users to color hex-cells according to their quality using one of three color-maps, shown in Fig.~\ref{fig:quality_filter_img}, bottom, that have been chosen for their readability and wide circulation.
The quality coloring works in addition to the filtering mechanism which isolates badly shaped elements by hiding all elements which are better than a prescribed threshold measure (see Sec.~\ref{sec:filters} and Figure \ref{fig:clipping_plane_img}, right).

We also display a histogram to show the distribution of the quality values among all the elements (Fig.~\ref{fig:quality_filter_img}, bottom right). The histogram, which consists of either vertical or horizontal bars (users' choice), can be saved as a subimage, it is color-coded using the currently selected color-map, and doubles as a labelled legend for the current color-coding on the 3D rendering.

\subsection{Supported quality metrics}
\label{sec:measures}
There is a wide spectrum of different metrics that can be useful to assess the quality and other characteristics of a hex-mesh.
Many of these metrics refer to the deviance from the \emph{ideal} shape, that is, a perfect cube having planar faces, orthogonal angles and all edges with equal length. Broadly speaking, the larger the deviation from the ideal shape, the more inaccurate the results produced by a FEA simulations can be expected. In particular, elements with nearly zero volume (i.e., \emph{degenerate}) or negative volume (i.e., \emph{inverted}) may introduce a significant error, or even preclude the entire simulation \cite{Livesu:2015}. For a discussion of these metrics, the reader is referred to the recent interesting study \cite{Gao2017Evaluation}.

\hexalab supports all the quality metrics reported in Verdict \cite{stimpson2007verdict}, that are the most widely adopted library in the field for the evaluation of finite elements. We report them in Table~\ref{tab:quality_metrics}.

\begin{figure*}[t]
	\centering
	\begin{tabular}{@{}c@{}c@{}c@{}}
		\includegraphics[width =0.34\linewidth]{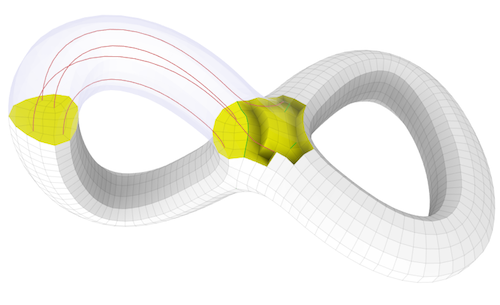}&
		\includegraphics[width =0.34\linewidth]{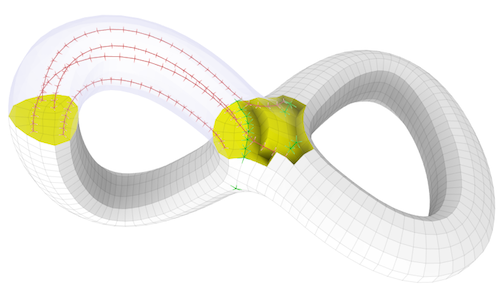}&
		\includegraphics[width =0.34\linewidth]{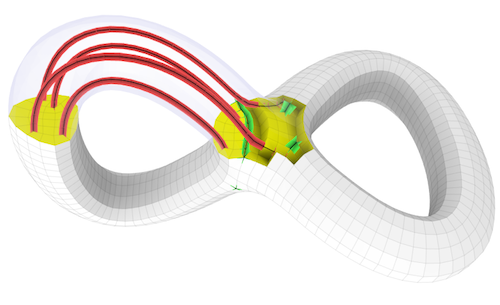}\\
		\includegraphics[width =0.34\linewidth]{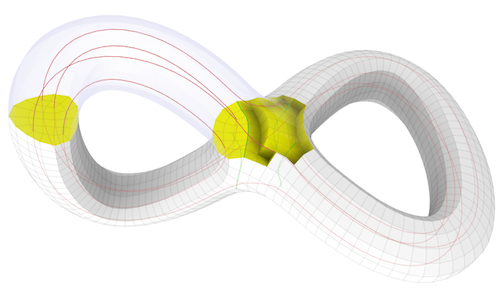}&
		\includegraphics[width =0.34\linewidth]{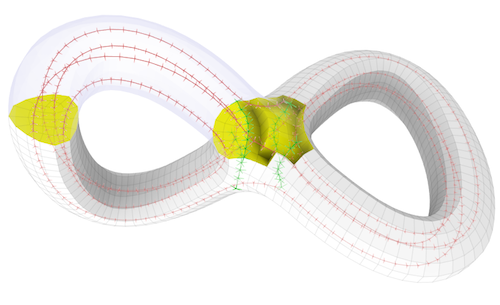}&
		\includegraphics[width =0.34\linewidth]{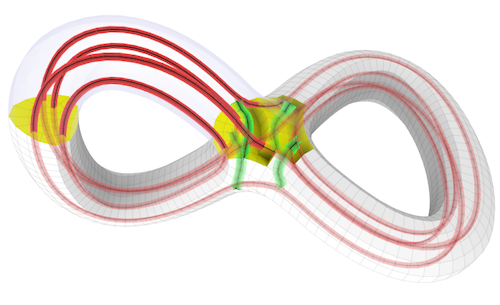}\\
	\end{tabular}
	\caption{The sequence of visualization modes for the irregular structures. In increasing order of amount of information provided:
	``wire'' mode (default), i.e. as lines colored as a function of edge arity; ``barbed wires'' mode, i.e.\ with adding all edges stemming out from the irregular edges; ``paper'' mode, where incident faces are also partially shown.
	In modalities variants shown on bottom, \hexalab reveals in transparency elements which would otherwise be occluded by currently un-filtered hexas. Model courtesy of \cite{Li:2012}.}
	\label{fig:irregular}
\end{figure*}

\subsection{Metric normalization}
The supported metrics differ widely in range and behavior; for example the range is bounded for some measure (for example, the \emph{Scaled Jacobian}) and unbounded for others (for example, cell \emph{volume}); optimal value can be the lowest or the highest value in the range.

In spite of this heterogeneity, in order to manage the metrics in a consistent yet intuitive way, \hexalab \emph{internally} stores per-cell element values in a normalized interval $[0,1]$, with the convention (where appropriate) that the value $0$ always corresponds to the worst quality, and the value $1$ to the best. To this end, we have designed, for each metric $i$, an ad-hoc function $\f_i$ which maps the original range of that metric into the normalized interval.  For metrics with unbounded ranges, $\f_i$ is also a function of the maximum and minimum values found in the current mesh (which are updated at load time). All these mapping functions are reported in Table~\ref{tab:quality_metrics}.

Thanks to this normalization, \hexalab will consistently color the worst elements (for example, as \emph{red} under ``Jet'' color map), will consistently isolate the worst elements with the quality filter, and will consistently account for the worst elements in the right (or bottom) end of the histograms, regardless of the currently selected measure.
The normalized values, however, are only used internally and never directly exposed to the user: \hexalab always uses the original values of the current metric on all histogram labels, legends, and GUI elements (e.g.\ the box containing the threshold value of the quality filters). This is well-defined as all our functions $\f_i$ are invertible (technically, with the exception below).

The Scaled Jacobian is one of the most widely adopted quality measures for 
hexmeshes, and is the default measure in \hexalab.
For this measure, which is defined in the range $[-1,1]$, $\f_i$ simply clamps negative values to 0. We consider this mapping more useful than a linear mapping from $[-1,1]$ into $[0,1]$.
First of all, $\f_i$ correctly assigns a very poor quality to cells with a Scaled Jacobian value close to zero. Also, cells with inverted (or degenerate) corners, which can be argued to be equally invalid in hexmeshes, are clustered in one histogram bin (zero); consequently,
the first bin of the histogram counts inverted (and degenerate) cells, and pushing the quality filter all the way to the right hides every cell but inverted (and degenerate) ones.

%% file: structure_vis.tex
\section{Mesh structure Visualization}
\label{sec:structure}


In a structured hex-mesh, internal vertices are shared by eight cells, and internal edges by four cells; boundary edges (edges lying on the boundary of a mesh) are shared by two cells, and boundary vertices by four. Elements (vertices or edges) with a mismatching number of adjacent cells are termed \emph{irregular}. 
%
%
Irregular edges are necessarily organized in strips which traverse the mesh from an irregular vertex to another (either on the surface or in the interior of the mesh). 

Similarly to irregular vertices on quad meshing, irregular elements on hex-meshes characterize how cells are globally organized.
For example, a particular class of hex-meshes, called \emph{generalized} polycubes \cite{generalizedPolycubes}, 
have no internal irregular edges. 
Internal irregular elements, often referred to as ``meshing singularities'', have been closely investigated for their direct relationship with the subdivision of hex-meshes into fully structured sub-blocks \cite{nonni,gao2017simplification}.
In general, the number, disposition and connection of irregular edges are linked to important properties of the hex-mesh.

\Hexalab offers a variety of visualization modalities to depict irregular elements. In each variant, different primitives are used to convey the presence of irregular edges, and (some of them) associated info like their valence. All the modalities are illustrated in Fig.~\ref{fig:irregular}. 

We offer a series of visualization modalities for irregular elements which range from lightly hinted to most informative (but also potentially cluttering and confusing the rest of the visualization). In the most informative modality, the irregular structure is visible in semi-transparency through the cell elements, whereas with the lighter modes it is only exposed in areas where cells are hidden by filters.
The user selects one modality through a slider having as many discrete steps as the number of possible choices. The leftmost step disables the irregular-element visualization entirely, while the rightmost step corresponds to the most informative modality. 




%% file: quality_measures.tex

%% file: status.tex

\section{Status snapshots}
\label{sec:status}
\begin{figure}[tbp]
	\centering
\includegraphics[width=0.45\linewidth]{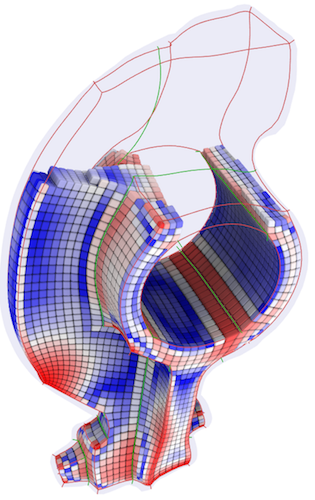}
\includegraphics[width=0.45\linewidth]{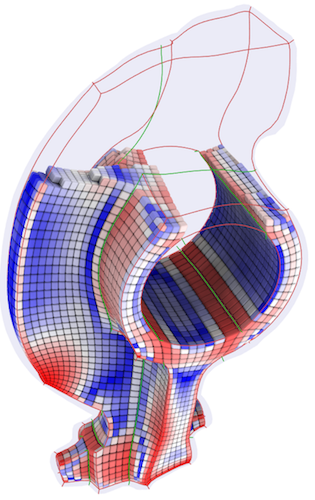}
	\caption{Two alternative meshings for the Rockerarm model 
	(in this instance, the input and output of the technique \cite{10.1111:cgf.12959})
	visually compared in HexaLab by using with the same \emph{status}. The status has been copy-pasted across visualization sessions.}
	\label{fig:status}
\end{figure}

The current ``status'' of the application, which includes current settings for all visualization tools and camera parameters, is unambiguously described in textual format, as JSON object. 
This includes, for example, the orientation and position of the slicing plane, the view direction, the used colors, and so on.
The main intended use is to ease direct comparisons between different hex-meshes  (see Fig.~\ref{fig:status}),
and it is also the base for a number of additional mechanisms:
\begin{description}
\item[Manual status copy-pasting:] the user can copy the status to and from the system clipboard (via the copy and paste keyboard shortcuts), and thus manually transfer the visualization settings across simultaneously open instances of HexaLab (or across consecutive visualization sessions, storing them for further reference);


\item[Image reproducibility:] the status is automatically attached as meta-data to every produced image (via a snapshot tool), for future reference. Dragging a snapshot of a previous session into \hexalab, meta-data will be read and the status of the web application automatically updated so as to reproduce the same visual contained in the image. All the images contained in this paper have been produced with this tool. Exporting images from the original version of the manuscript and dragging them into \hexalab will allow  readers to see the same images in their browsers;

\item[Manual parameter tweaking:] as the setup is saved on a text file, the user can manually edit every parameter. While not intended as the main way to interact with HexaLab, this allows for full control (e.g.\ to pinpoint the orientation of the cutting plane, to select certain cells by numeric ID, to choose colors). Secondarily, it partially exempts the graphic interface from the need to provide full  access to each parameter, allowing us to simplify it;
\end{description}


%% file: impl.tex

\section{Implementation}

The software architecture of \hexalab loosely follows the Model-View-Controller design pattern \cite{Leff:2001}. Two modules are integrated in a single web application: a back-end module that deals with mesh input/output, storage, and manipulation, and a front-end module for the rendering and the GUI.

\subsection{Internal hex-mesh representation}
Different representations for polyhedral meshes offer several tradeoffs between navigation efficiency and memory footprint. Our requirements include the necessity to keep the latency time low when the current filtering changes and a new quad-mesh boundary of visible elements needs to be produced. 
We rely on assumption that the mesh connectivity is fixed, and that the mesh resolution is in the range of around $10^5$ cell elements.
 
We opted to store the hex-mesh \meshv as a \emph{generalized map}. One reason is that homogeneous array sizes simplify algorithms; also, this makes \hexalab amenable to include support of non hexahedral polyhedra in future expansions. 

Generalized maps are a slightly more general but less compact version of a combinatorial map; \hexalab uses its own implementation, although other implementations are available \cite{cgal_cmaps,suggestedByRev3:2013}. They are formalized in \cite{Lienhardt91,Lienhardt94}, and here we only briefly recap them. The mesh connectivity is stored as a collection of ``darts'', which are a generalization of the half-edge structure commonly used for polygonal meshes.
A dart represents a topological location on the mesh, and consists of a collection of four indices to one 3D, 2D, 1D, and 0D element of the mesh (i.e.
a cell, a face, an edge, and a vertex). Additionally, each dart stores a set of four indices (called ``involutions'') pointing to the dart reached if any of its four elements is swapped (invalid indices are stored at darts at mesh boundaries). Involutions provide an efficient mean to navigate over the mesh. We precompute and store darts immediately after mesh import, leveraging standard associative container structures (such as black-red trees). Finally, we store attributes at mesh elements, such as $(x,y,z)$ positions at vertices, normal vectors at faces, and ``filtered-status'' (a Boolean variable) at cells.

\subsection{Rendering algorithms}
\label{ambientocclusion}
The extracted quad-mesh \meshs is stored as an indexed triangular mesh, duplicating vertices when we have to represent normal or color discontinuities along edges. Quads are simply split into triangle pairs along an arbitrary diagonal (although rendering methods which bypass the need for this arbitrary split have been proposed, \cite{hormann2004quadrilateral}).

By default, Ambient Occlusion (AO) is the only illumination component used in \hexalab and substitutes direct illumination entirely. 
Similarly to \cite{Tarini2006}, AO terms are iteratively computed in object space, by superimposing a sequence of shadow-maps, one for each probe light direction: in our case, we accumulate unblocked light, weighted by the \emph{Cosine law}, directly at vertices of \meshs; note that this potentially produces discontinuity of ambient occlusion factors at normal discontinuities, as expected. We use a set of 1024 probe lights directions, which sample the unit sphere in an approximately uniform way, and are randomly constructed  approximating a blue noise distribution.
All renderings, including of the shadow-map, and the accumulate light on the per-vertex AO terms, are performed in GPU, leveraging WebGL. 

AO terms are updated in background, accumulating contributions from light probes without compromising interactivity. 
After any user-triggered update of \meshs, buffered AO terms are discarded, and their computation restarted.
We wait that at least six light probes are accumulated in the AO terms before using them in any rendering;
before then, we fall back screen-space AO approximation \cite{Bavoil2008,Mittring2007,Ritschel2009}.





\subsection{Polygonal mesh extraction}
\label{sec:round_mesh}

After each user filter operation, the hidden cells are removed, then \Hexalab dynamically extracts a polygonal surface-mesh \meshs to be globally lit and rendered from 
\meshv.

In \emph{darkness-coded edges} mode,  \meshs is simply composed of the quad faces of \meshv separating one filtered and one unfiltered cell of \meshv.
In \emph{fissure}  and \emph{rounded} mode, visible cells undergo certain spatial deformations, affecting both the polygon connectivity and the geometry of \meshs, and requiring specialized algorithms to extract it. Specifically, in \emph{fissure} mode, cells are shrunk, forming gaps around quads of \meshv; in \emph{rounded} mode, cells edges are rounded, forming, as a side effect, small tubular gaps around edges of \meshv. 

For the latter two modes, our mesh extraction algorithm consists of two steps: first we identify the vertices of \meshv which lie on its boundary (given the current filtering), and we label them as \emph{exposed}; next, we process any cell with at least one exposed vertex, and we add vertices and polygons to \meshs according to the exposed status of the eight corner vertices of that cell, as detailed in Sec.~\ref{sec:fissureExtr} and \ref{sec:roundedExtr}.

\begin{figure}[tbp]
	\centering
\includegraphics[width =0.25\linewidth]{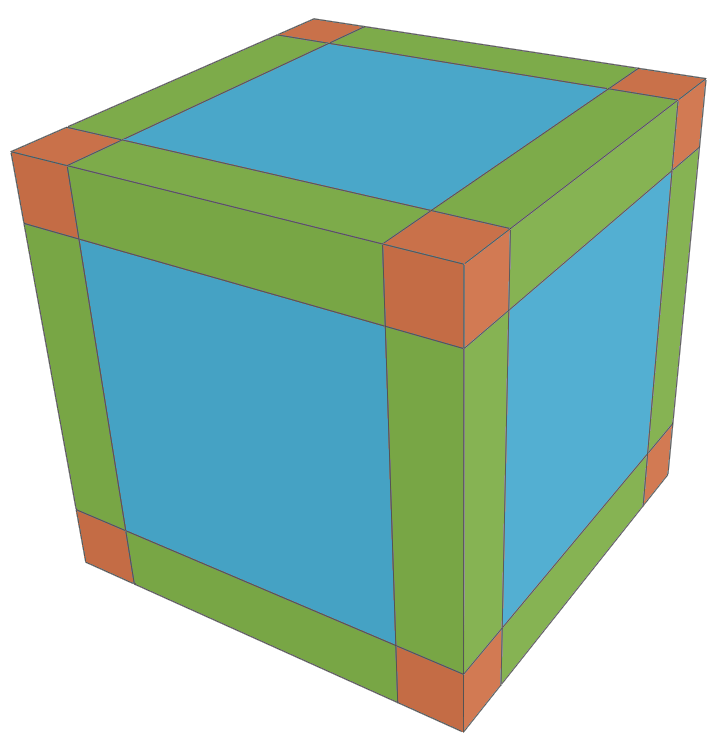}%
\includegraphics[width =0.25\linewidth]{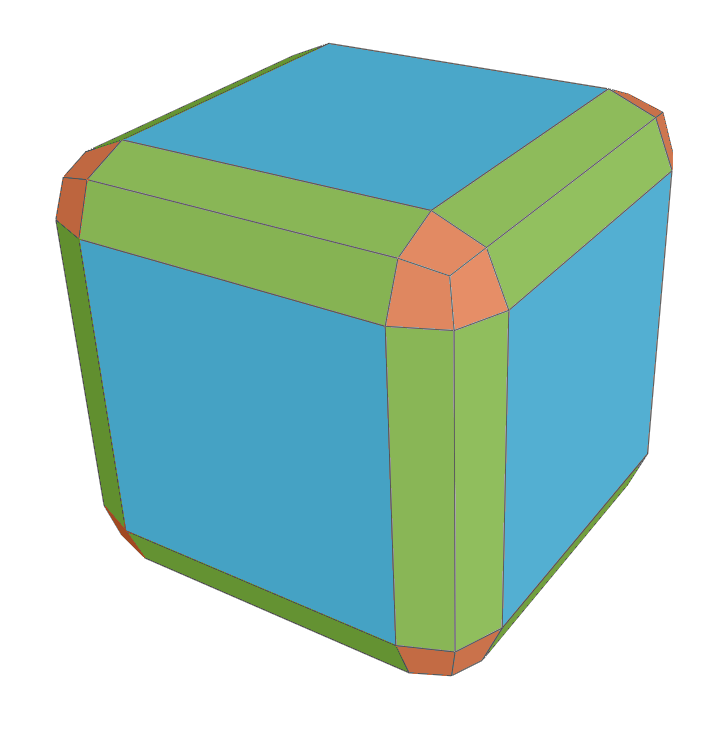}\\
\includegraphics[width =0.25\linewidth]{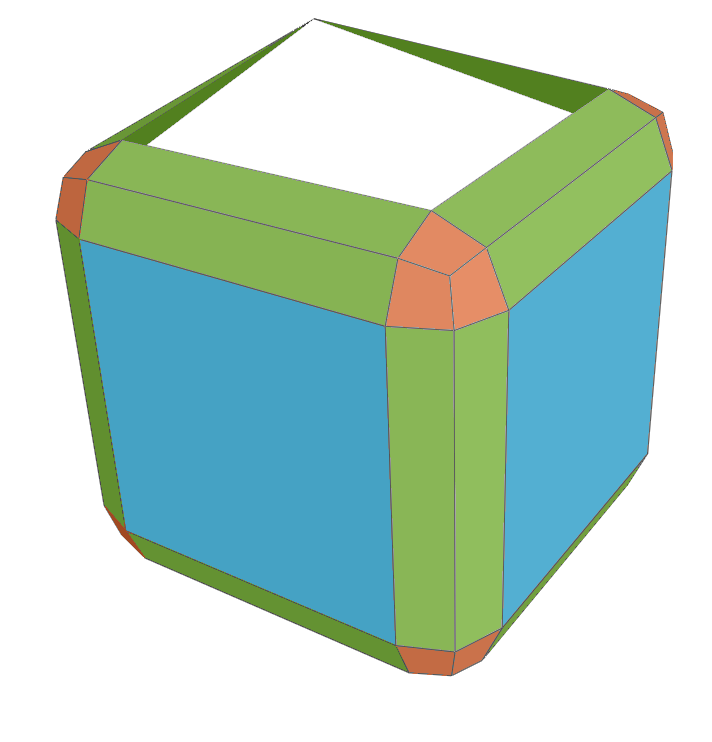}%
\includegraphics[width =0.25\linewidth]{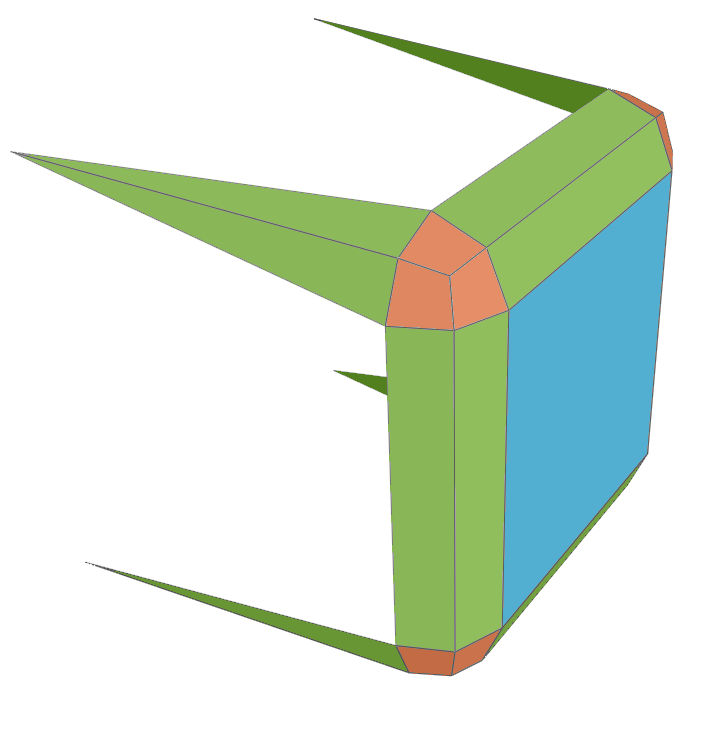}%
\includegraphics[width =0.25\linewidth]{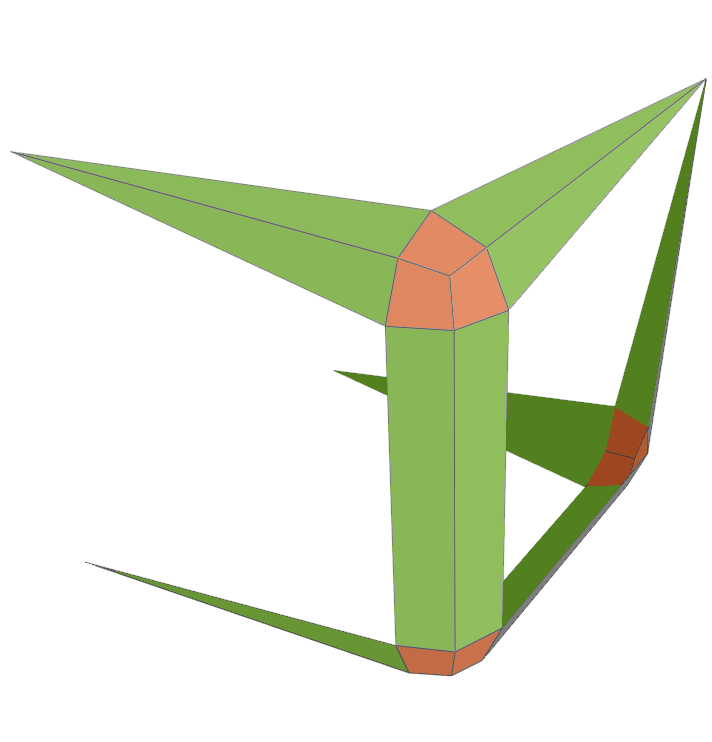}%
\includegraphics[width =0.25\linewidth]{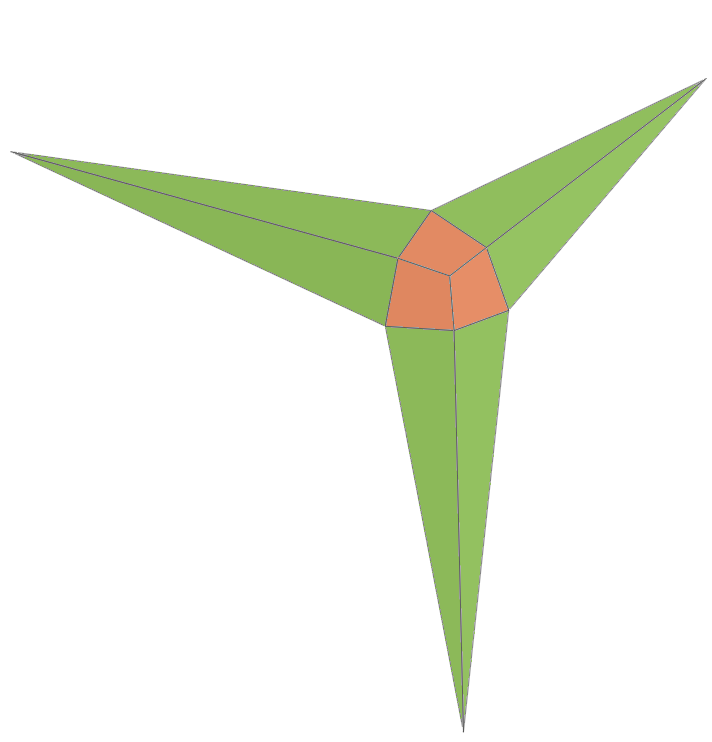}%
\caption{Boundary polygons produced for a hexacell (to be rendered and lit). Top: topological subdivision and geometrical displacement.
	Bottom: examples of the produced polygons for four different configurations of six, four, three, and one ``exposed'' vertices.}
	\label{fig:rounded}
\end{figure}

In all modes, \meshs is always a closed and geometrically watertight polygonal mesh.
Its vertices are produced with positions, base color, and normals (used by the global lighting).
In \emph{darkness-coded edges} and \emph{fissure} mode, we employ flat shading, so each quad of \meshs indexes its own instances of the vertices (producing hard creases between all neighboring faces). In \emph{rounded} mode, we employ smooth shading, and polygons of \meshs belonging to the same cell index share vertices (so hard creases only appear \emph{between} cells). 

\subsubsection{Mesh extraction in Fissure mode}
\label{sec:fissureExtr}
We produce a quad for each cell face sharing at least one exposed vertex (producing all four \meshs vertices for that face, irrespective of the ``exposed'' status of the corresponding \meshv vertex).  Exposed vertices only are then moved toward the barycenter of the cell, covering a small, user-selected percentage of the distance. This way, inter-cell gaps are formed, but only in the proximity of the boundary of the visible parts of \meshv; because the unexposed vertices are kept in their original positions, the gaps close going toward the internal portions of the hex-mesh, and \meshs has no element at all in the more internal parts.  

This introduces an approximation, as in reality every side of every cell of \meshv would be exposed due to all fissures forming one connected empty space. 
The visual effect of the approximation is, however, extremely small, because it takes place in regions which are typically both occluded by more superficial elements and made dark by the global lighting. 
The approximation drastically improves performances.


Because vertices are moved toward the barycenter of the respective cell covering a fixed proportion of the distance, the schema automatically adapts to varying sizes of the cells (e.g.\ narrower gaps are formed around smaller cells).

\subsubsection{Mesh extraction in Rounded mode}
\label{sec:roundedExtr}
In this mode, we subdivide each side of the processed cell into $3\times 3$ squared sub-faces; this produces four additional vertices at each cell face, and two additional vertices at each cell edge (see Fig.~\ref{fig:rounded}). The vertices on the edges and corners of the cell are displaced toward the barycenter of the respective edges to produce the actual rounding (similarly to the previous case, this makes the size of the roundings to automatically adapt to the local size and shape of the processed cell).

A \emph{side} face (blue in Fig.~\ref{fig:rounded}) is produced only when all the four corresponding \meshv vertices are exposed. 
A set of three \emph{corner} faces (red in Fig.~\ref{fig:rounded}) is only produced when the corresponding cell vertex is exposed. 
A set of two \emph{edge} faces (green in Fig.~\ref{fig:rounded})  is only produced if either or both the vertices on the corresponding cell edge are exposed;
when only one is exposed, then the two faces are reduced to triangles, and the unexposed vertex is not displaced and kept to the original vertex position
(Fig.~\ref{fig:rounded}, bottom).
In this way, the tubular gaps around edges are artificially closed going toward the interior parts, leaving \meshs geometrically watertight. 

Similarly to the fissure mode, we willingly introduce an approximation to improve rendering performances. The visual impact of such an approximation is completely or almost completely negated both by occlusions and by lack of light reaching the affected parts.


A vertex is added to \meshs only if it belongs to at least one produced face. The vertices of the side face (blue in Fig.~\ref{fig:rounded}) are assigned constant normals, making that face appear visually flat. The normal interpolation is limited to edges and corner faces (green and red in Fig.~\ref{fig:rounded}).

\subsection{Supported file formats}
\Hexalab supports two among the most widely used file formats for the exchange of hex-meshes, namely the MEDIT format \cite{MEDIT} (.mesh filename extension) and the VTK library format \cite{VTK} (.vtk filename extension). Currently, \hexalab importer parses only the hexahedral elements expressed in these two formats, ignoring any other polyhedra (e.g., tetrahedra, pyramids and wedges). 
%

\subsection{Batch processing}
\Hexalab can load a zipped archive containing a collection of hex-meshes, and produces a zipped archive containing one screenshot (and one quality histogram) for each model, all sharing the same settings.

\subsection{Employed Tools}
The back-end, that deals with mesh input/output and geometric and topologic analysis, is developed in \emph{C++} using \emph{Eigen} library \cite{eigenweb} for the linear algebra computations; for allowing the execution of C++ code on the browser client we used \emph{Emscripten} \cite{emscripten} to ``transpile'' it in \emph{asm.js} (a low-level subset of JavaScript) that is recognized and efficiently executed by modern js engines. The front-end, shown in Fig.~\ref{fig:interface}, is developed directly in JavaScript, using standard web tools (HTML 5.0, CSS, AJAX, and \emph{jQuery}) for the GUI, \emph{webGL} and \emph{Three.js} for the rendering and the trackball, and \emph{Plotly.js} \cite{plotly} for the graph plots. The code has no other dependencies. We use \emph{GitHub} as an open-source repository for the code and the meshes and, thanks to the fact that it is a pure client web application, also for the web hosting.

%% file: end.tex

\section{Conclusions}
We presented a novel tool for interactive hexahedral mesh visualization and first analysis, which combines a number of highly customizable tools to explore different features and characteristics of the inspected mesh. It produces readable images which convey both shape, quality, and the topological structure of the inspected hex-meshes, including its internal parts, plus simple numeric measures in form of graphs and data (according to a number of widely accepted measures). To this end, also \hexalab introduces several new visualization modalities specifically designed for this purpose, and employs a real-time global illumination model.
It also provides direct access to a repository of results from several recent State-of-the-Art hex-mesh creation and processing solutions, thus easing further research by providing an easy way to compare against them. All data and images produced with \hexalab are easily reproducible and can be streamlined, specifically all the images shown in this paper can be re-created by loading into \Hexalab the appropriate model and dragging the png file of each figure over the application window. 

The tool is immediately available to researchers and practitioners in the form of an easily accessible 3D web-application (which is cross-platform, cross-browser, and cross-vendor, and requires no installation), and as an Open-Source project. 
We expect that it can be employed, in research, as a tool to gain insights on hex-meshes and therefore on the algorithms to produce, manipulate, and process them, and also to help the dissemination of engineering and scientific results (by producing high quality images fit for scientific articles and presentations).



\subsection{Current limitations and future work}


As mentioned in Sec.~\ref{sec:repository}, one main planned activity consists in keeping the repository of State-of-The-Art hex-meshes up-to-date. 
There are several directions in which \hexalab can be expanded.

\noindent\textbf{Generalization to polyhedral meshes:} currently, \hexalab only supports pure hex-meshes. Other important classes of polyhedral meshes include (pure) tetrahedral meshes, and hybrid hex-dominant meshes. These classes are the focus of recent research works (e.g. \cite{tetwild} and \cite{hybridHexa} respectively). Many of the mechanisms employed by \hexalab are, in principle, extendible to them, including the filtering tools, the visualization modes, the internal data structures. Quality measures would have to be adapted, and it is not clear how to do so for hex-dominant meshes.

\noindent\textbf{GUI improvements:} in this project, we did not focus on GUI design, although preliminary testing with fellow researchers indicates that our GUI performs satisfactorily for our purposes. A deeper design analysis, including user studies, can be performed in the future to provide \hexalab with an improved interface. Note that this tool is intended for practitioners (modelers, architects, structural engineers, researchers on fields such as geometric processing and 3D computer graphics) and therefore its GUI targets expert users.

\noindent\textbf{Improved visualization of the global structure:} currently, Hexa-Lab highlights irregular edges, but additional mechanisms could be used to communicate the global structure of the inspected hex-mesh. For example, future versions could compute and use the base complex, or integrate the advanced visualization methods proposed in \cite{newCitation}, which target the visualization of the (potentially intricate) whole structure of a hex-mesh.

\noindent\textbf{Dynamic datasets:} currently, \Hexalab only supports static meshes. There is a research interest on dynamic datasets, which could be for example provided for visualization as different key-frame meshes sharing the same connectivity but different geometry, to be interpolated. 


\noindent\textbf{Attributes:} \Hexalab currently focuses on the shape of the mesh only, and its visualization disregards attributes which are often defined on its elements (such as vertices, or cells). 

\noindent\textbf{Trackball:} \Hexalab currently employs a standard sphere-based trackball to let the user orient the mesh, but a ``generalized trackball'' \cite{generalizedTrackball2016} could be employed in its place.
